\documentclass[english,prb,aps,amsmath,amssymb,reprint,superscriptaddress,showpacs]{revtex4-2}
\usepackage[T1]{fontenc}
\usepackage[utf8]{inputenc}
\usepackage{color}
\usepackage{babel}
\usepackage{array}
\usepackage{textcomp}
\usepackage{multirow}
\usepackage{amsmath}
\usepackage{amssymb}
\usepackage{graphicx}
\usepackage[unicode=true,
 bookmarks=true,bookmarksnumbered=false,bookmarksopen=false,
 breaklinks=true,pdfborder={0 0 0},pdfborderstyle={},backref=false,colorlinks=true]
 {hyperref}
\hypersetup{pdftitle={Crucial role of vibrational entropy on the Si(111)-7x7 surface stability},
 pdfauthor={R. Zhachuk, J. Coutinho},
 pdfsubject={68.35.bg, 68.35.Gy, 68.35.Md},
 pdfkeywords={Silicon, Germanium, Surface strain, Surface stress},
 allcolors=blue}

\makeatletter

\newcommand{\lyxmathsym}[1]{\ifmmode\begingroup\def\b@ld{bold}
  \text{\ifx\math@version\b@ld\bfseries\fi#1}\endgroup\else#1\fi}

\providecommand{\tabularnewline}{\\}

\makeatother

\begin{document}
\title{Crucial role of vibrational entropy in the Si$(111)$-$7\times7$
surface structure stability}
\author{R. A. Zhachuk}
\email{zhachuk@gmail.com}

\affiliation{Institute of Semiconductor Physics, pr. Lavrentyeva 13, Novosibirsk
630090, Russia}
\author{J. Coutinho}
\affiliation{I3N, Department of Physics, University of Aveiro, Campus Santiago,
3810-193 Aveiro, Portugal}
\date{\today}
\begin{abstract}
We investigate the relative thermodynamic stability of the $3\times3$,
$5\times5$, $7\times7$, $9\times9$ and infinitely large structures
related to the dimers-adatoms-stacking faults family of Si$(111)$
surface reconstructions by means of first-principles calculations.
Upon accounting for the vibrational contribution to the surface free
energy, we find that the $5\times5$ structure is more stable than
the $7\times7$ at low temperatures. While a phase transition is anticipated
to occur at around room temperature, the $7\times7\rightarrow5\times5$
transformation upon cooling is hindered by the limited mobility of
Si atoms. The results not only flag a crucial role of vibrational
entropy in the formation of the $7\times7$ structure at elevated
temperatures, but also point for its metastable nature below room
temperature. \emph{{[}Pre-print published in Physical Review B }\textbf{\emph{105}}\emph{,
245306 (2022){]}} \href{https://doi.org/10.1103/PhysRevB.105.245306}{DOI:10.1103/PhysRevB.105.245306}
\end{abstract}
\maketitle

\section{Introduction}

First-principles electronic structure methods, most notably density
functional theory (DFT), have been widely used as a powerful tool
to solve surface structures and to study their electronic properties.
Due to its technological importance, semiconductor silicon has been
a benchmark for studying surface physics, with several surface structures
being extensively investigated \citep{dab94,erw96,bech01,bat09,punkkinen14,zha17,zha19,zha20,zha20-2,zha21}.
When Si surface structures are composed of different building blocks
as, for example, $2\times1$, $2\times2$ and $7\times7$ reconstructions
on Si$(111)$, their surface energy differences are usually significant
(above $1\ \mathrm{meV/\mathring{A}^{2}}$), facilitating a reliable
identification of the lowest energy atomistic configuration \citep{ste02,zha13}.
In the above, the $2\times1$ structure consists of $\pi$-bonded
chains \citep{pan81}, $2\times2$ consists of Si adatoms on $\mathrm{T_{4}}$-sites
\citep{zha17-2}, while the lowest energy $7\times7$ reconstruction
is described by the complex dimers adatoms stacking faults (DAS) model
\citep{tak85}. The DAS model describes a series of structures belonging
to the $\left(2n+1\right)\times\left(2n+1\right)$ family of reconstructions
observed on Si$(111)$ and Ge$(111)$ surfaces ($n$ being a positive
integer): $3\times3$, $5\times5$, $7\times7$, $9\times9$ and so
on.

However, if the competing surface structures belong to the same family
of reconstructions, \emph{i.e.} when they are made of common building
blocks, it becomes difficult to determine which one has the lowest
formation energy, and ultimately to find the one that should be observed.
For instance, the $c(4\times2)$ and $p(2\times2$) reconstructions
on Si$(100)$ are both composed of Si dimers, but because they buckle
with different periodicity, their reported energy difference is as
low as $0.7\,\mathrm{meV/dimer}$ ($0.02\ \mathrm{meV/\mathring{A}^{2}}$)
\citep{sei04}. We note that in spite of such low energy difference,
the calculated lowest energy structure {[}$c(4\times2)${]} is in
agreement with experimental scanning tunneling microscopy (STM) data
obtained at low temperature.

The $5\times5$ and $7\times7$ DAS reconstructions on Si$(111)$
surface provide us with another example of two structures with very
close surface formation energies. Two previous attempts to compare
the surface formation energies of DAS reconstructions on Si$(111)$
within DFT are singled out. In Ref.~\onlinecite{sti92}, formation
energies of $3\times3$, $5\times5$ and $7\times7$ reconstructed
Si$(111)$ surfaces were compared, and the $7\times7$ reconstruction
was claimed as the most favorable structure. This result agrees with
many experiments demonstrating that the $7\times7$ is the most frequently
observed structure on Si$(111)$ below the $7\times7\leftrightarrow1\times1$
order-disorder transition temperature at around $1100\,\mathrm{K}$
\citep{ish91}. However, two years later, Needels \citep{nee93} provided
evidence that the above calculations suffered from poor Brillouin
zone (BZ) sampling, that once adequately corrected, resulted in a
different energy ordering, with the $5\times5$ surface becoming the
ground state (more stable than the $7\times7$ reconstruction).

The second attempt to calculate the formation energy of Si$(111)$
DAS reconstructions was carried out by Solares \emph{et\ al.} \citep{sol05}.
The Si$(111)\textrm{-}7\times7$ came out as the one with lowest formation
energy, and once again, was deemed the most stable. However, the calculated
energy difference between $5\times5$ and $7\times7$ structures of
only $4\ \mathrm{meV/1\times1\ cell}$ ($0.3\ \mathrm{meV/\mathring{A}^{2}}$),
which combined with a reported accuracy of about $28\ \mathrm{meV/1\times1\ cell}$
($2.2\ \mathrm{meV/\mathring{A}^{2}}$) \citep{sol05}, make their
conclusions questionable. The large error stems from several calculation
settings, most notably the use of thin 4-bilayer (BL) slabs, insufficient
number of $\mathbf{k}$-points to sample the BZ, and the use of the
experimental lattice parameter ($5.43\,\mathrm{\mathring{A}}$) instead
of calculated one ($5.48\,\mathrm{\mathring{A}}$) in the construction
of the models \citep{sol05}. The later factor introduces an artificial
strain field across the slabs which alter the calculated energy differences
of $5\times5$ and $7\times7$ surface structures.

In summary, the available picture regarding the stability ordering
of the DAS structures on Si$(111)$ is rather unclear. In particular,
our understanding of the physics that governs the formation of the
most stable Si surface is covered by error bars. Importantly, both
studies described above are based on calculations of potential (static)
energies, \emph{i.e.} for $0\ \mathrm{K}$ temperature only. It is
a general practice to neglect thermal excitations as well as the contribution
of entropy to the free energy of formation of surfaces. However, given
the small energy difference between $5\times5$ and $7\times7$ surface
structures, it would be important not only to account for the above
referred sources of uncertainty (slab thickness, BZ sampling, artificial
strain neutralization), but also to assess the effect of entropy,
especially the contribution of vibrational degrees of freedom, usually
the dominant finite-temperature contribution to the surface free energy
of formation of semiconductor surfaces.

The aim of the present work is to find well converged energy differences
between several members of the DAS family of structures on Si$(111)$
surface, size the entropy contribution to the free energy of formation
of $5\times5$ and $7\times7$ reconstructions, and finally understand
the driving stabilization factors. We find that while the $5\times5$
reconstruction is the most favorable structure at $T=0$\ K, vibrational
freedom stabilizes the $7\times7$ reconstruction, with a cross-over
of the respective free-energies of formation taking place close to
room temperature. The observation of metastable Si$(111)$-$7\times7$
below room temperature can be explained by a large barrier along the
$7\times7\rightarrow5\times5$ transition that involves the motion
of Si atoms during cooling.

\section{Calculation details, Results and Discussion}

\subsection{Relative formation energies of Si(111) surfaces with DAS reconstructions
at $0\,\mathrm{K}$ temperature}

The calculations were carried out using the pseudopotential \citep{tro91}
density functional theory \texttt{SIESTA} code \citep{sol02,gar20}.
Both the local density approximation (LDA) of Perdew and Zunger (PZ)
\citep{per81}, as well as the generalized gradient approximation
(GGA) of Perdew \emph{et\ al.} (commonly referred as PBEsol) \citep{per08},
were employed for the description of the exchange-correlation (XC)
interactions between electrons. Although it was previously shown that
DAS Si$(111)$ surfaces are diamagnetic \citep{sol05}, our calculations
were spin-unrestricted. The valence states were expressed as linear
combinations of Sankey-Niklewski-type numerical atomic orbitals \citep{sol02}.
In the present calculations, polarized double-$\zeta$ functions were
placed on all atomic coordinates. This means two sets of \emph{s}-
and \emph{p}-orbitals plus one set of \emph{d}-orbitals on silicon
atoms, and two sets of \emph{s}-orbitals plus one set of \emph{p}-orbitals
on hydrogen atoms. The electron density and potential terms were calculated
on a real space grid with spacing equivalent to a plane-wave cut-off
of 200 Ry.

\noindent 
\begin{figure}
\includegraphics[clip,width=8.5cm]{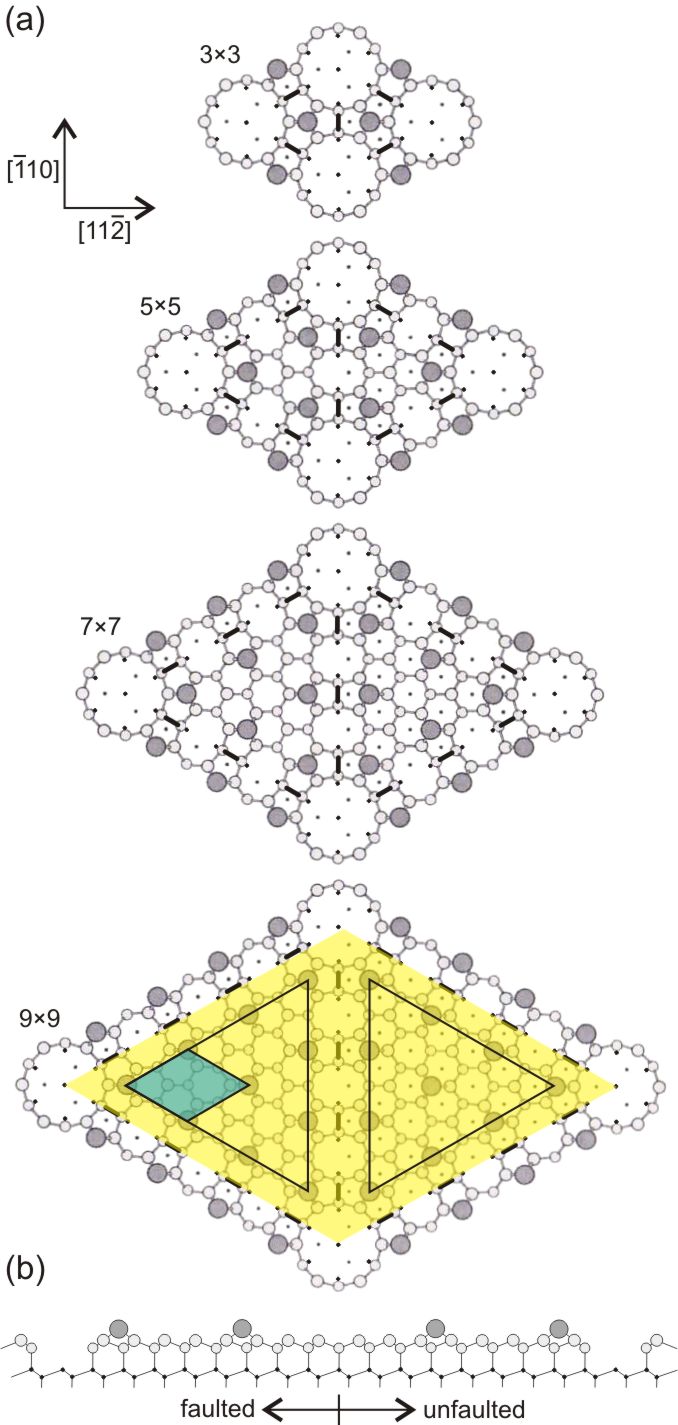}

\caption{\label{fig1} (a) Top view of $3\times3$, $5\times5$, $7\times7$,
and $9\times9$ structures on Si(111) surface according to the DAS
model. Dark-shaded large circles are adatoms. Light-shaded small circles
are atoms in the upper bilayer and in the surface dimes. Black dots
are atoms in the bulk unreconstructed bilayer. Thick rods edging the
perimeter of the cells are dimers. For the $9\times9$ structure the
unit cell is delimited by a yellow rhombus, triangles outline the
$2\times2$ domains in faulted and unfaulted half unit cells, and
the blue rhombus shows a $2\times2$ unit cell within a faulted half.
(b) Sectional view in the $(1\bar{1}0)$ plane cutting the long diagonal
of the $9\times9$ surface unit cell, showing the nearest-neighbor
bonding structure.}
\end{figure}

For total energy calculations we used Si$(111)$-slabs of 6~BLs separated
by a 30~Å thick vacuum layer. Dangling bonds at the \emph{bottom
side} of the slabs were saturated by hydrogen, while the top layers
were modified according to the DAS $3\times3$, $5\times5$, $7\times7$
and $9\times9$ atomic models \citep{tak85}, as well as a $2\times2$
model, consisting of adatoms on $T_{4}$ sites. A graphical description
of the surface structures is presented in Figures~\ref{fig1}(a)
(top-view) and \ref{fig1}(b) (cross-sectional view along the $[1\bar{1}0]$
direction).

The total number of Si atoms in the slabs were the following: 106
for $3\times3$, 300 for $5\times5$, 592 for $7\times7$, 982 for
$9\times9$ and 49 for $2\times2$. The positions of all slab atoms
(except for the H atoms and Si atoms in the bottom bilayer) were fully
optimized until the atomic forces became less than 1\ meV/Å. BZ integration
was discretized as summations over Monkhorst-Pack (MP) $\mathbf{k}$-point
meshes \citep{mon76}. The following BZ-sampling schemes were applied:
MP-$6\times6\times1$ for $3\times3$, MP-$4\times4\times1$ for $5\times5$,
MP-$3\times3\times1$ for $7\times7$, MP-$2\times2\times1$ for $9\times9$
and MP-$10\times10\times1$ for $2\times2$ surface slabs. These sampling
schemes lead to approximately the same $\mathbf{k}$-point density
in all BZs. We found that the above calculation settings provide a
relative surface energy convergence within $0.1\,\mathrm{meV/\mathring{A}^{2}}$.

Although experiments are normally performed under constant pressure,
any theoretical description of surface thermodynamics is more conveniently
carried out under constant volume. That avoids dealing with anharmonic
effects and the cumbersome connection between temperature and volume
via thermal expansion coefficients. As demonstrated by Estreicher
\emph{et\ al.} \citep{san03,est04}, the small thermal expansion
of crystalline Si leads to comparable constant-volume and constant-pressure
free energies up to a few hundred degrees Celsius. This is enough
for the present purpose. However, at higher temperatures, anharmonicity
and electronic excitations become important and the results should
be considered as qualitative.

The Helmholtz free energy (per unit area) of an infinite and periodic
semiconductor surface must account for electronic and vibrational
contributions, $f(T)=f_{\textrm{elec}}(T)+f_{\textrm{vib}}(T)$. Magnetic
degrees of freedom are not relevant since all structures considered
are diamagnetic. Rotational motion and configurational entropy are
also excluded for obvious reasons. The magnitude of the electronic
term ($f_{\textrm{elec}}$) is dominant, although it has a weak $T$-dependence.
On the other hand, the vibrational term accounts for a small zero-point
energy (ZPE) at $T=0$~K, but shows a relatively stronger $T$-dependence.

Mostly often, $f_{\textrm{elec}}$ is replaced by the internal electronic
surface energy at absolute zero temperature, $u_{\textrm{elec}}(T=0\,\textrm{K})=\gamma$,
as found from the stationary solution of the many-body electronic
problem within DFT. On the other hand, $f_{\textrm{vib}}$ is usually
neglected, making $f(T)=\gamma$, where for the sake of convenience
the electronic potential energy $\gamma$ is hereafter referred to
as surface energy. This approximation is well justified in many cases,
since $\gamma$ is usually much larger than the other terms. However,
as it will be shown below, this approach has been hiding an important
driving force that determines the structure of the Si(111) surface.

We are interested in finding relative surface free energies ($\Delta f$)
of Si(111) reconstructions. For that we will consider the Si$(111)\textrm{-}5\times5$
(DAS model) as reference. The energy of a Si$(111)\textrm{-}N\times N$
surface relative to that of Si$(111)\textrm{-}5\times5$ ($\Delta f_{N\times N}=f_{N\times N}-f_{5\times5}$)
is obtained from

\begin{equation}
\Delta f_{N\times N}=\frac{5^{2}F_{N\times N}-N^{2}F_{5\times5}-\mu\left(5^{2}X_{N\times N}-N^{2}X_{5\times5}\right)}{N^{2}5^{2}S_{1\times1}},\label{eq:delta_gamma}
\end{equation}
where to first approximation we start by assuming that $\Delta f_{N\times N}=\Delta\gamma_{N\times N}$,
the slab free energy $F_{N\times N}=E_{N\times N}$ with $E_{N\times N}$
being the total electronic energy of a particular $N\times N$ relaxed
slab made of $X_{N\times N}$ Si atoms ($N=$3, 5, 7 or 9), and $S_{1\times1}$
the area of the Si(111)-$1\times1$ surface unit cell. Analogously,
the Si chemical potential is approximated to the internal electronic
energy per Si atom in the bulk at $T=0$\ K, $\mu=\mu_{0}$, accounting
for any stoichiometric mismatch between $N\times N$ and $5\times5$
slabs. This quantity was found from a $1\times1$ bulk-like slab of
3 BLs thickness with MP-$20\times20\times8$ for BZ sampling.

In order to estimate the error bar due to different implementations
of DFT, including basis, pseudopotentials and exchange-correlation
functionals, we also calculated $\gamma$ for Si$(111)\textrm{-}7\times7$
and Si$(111)\textrm{-}5\times5$ using the \texttt{VASP} code \citep{kre96,kre96-2}.
These calculations were carried out within the GGA to spin-unrestricted
density functional theory, both without and with a correction for
dispersion forces (PBE \citep{per96} and PBE-D3 \citep{gri10,gri11},
respectively). Both \texttt{VASP} and \texttt{SIESTA} calculations
used identical conditions, except that in the former case, being a
planewave code, the Kohn-Sham states were described with planewaves
with kinetic energy up to 250\ eV.

\noindent 
\begin{table*}
\begin{ruledtabular}
\caption{\label{tab1}Relative surface energies ($\Delta\gamma_{N\times N}$)
of DAS-reconstructed Si$(111)$ surfaces ($\mathrm{meV/\mathring{A}^{2}}$)
calculated according to Eq.~\ref{eq:delta_gamma} from first principles,
using local basis and planewave DFT software packages, and several
XC functionals. The energy of the Si$(111)\textrm{-}5\times5$ surface
was taken as reference.}
\begin{tabular}{ccccccc}
\multirow{2}{*}{DFT software} & \multirow{2}{*}{XC functional} & \multicolumn{5}{c}{$\Delta\gamma_{N\times N}$}\tabularnewline
 &  & $3\times3$ & $5\times5$ & $7\times7$ & $9\times9$ & infinity\tabularnewline
\hline 
\multirow{2}{*}{SIESTA \citep{sol02}} & LDA (PZ \citep{per81}) & 2.58 & 0 & 0.15 & 0.57 & 3.95\tabularnewline
 & GGA (PBEsol \citep{per08}) & 2.32 & 0 & 0.23 & 0.7 & 4.16\tabularnewline
\hline 
\multirow{2}{*}{VASP \citep{kre96,kre96-2}} & GGA (PBE \citep{per96}) & - & 0 & 0.31 & - & -\tabularnewline
 & GGA (PBE-D3 \citep{gri10,gri11}) & - & 0 & 0.36 & - & -\tabularnewline
\end{tabular}
\end{ruledtabular}

\end{table*}

It is possible to calculate the formation energy of an infinitely
large DAS surface. To do this, we note that the upper bilayer of DAS
unit cell consists of two halves: one of them with and the other without
stacking faults (faulted and unfaulted halves in short), as represented
in Fig.~\ref{fig1}. These halves are linked together by dimers on
the surface \citep{tak85}. They also contain adatoms on top, arranged
according to a $2\times2$ periodicity. With increasing the size of
the DAS unit cell, the contribution of dimers to the surface energy
scales with $N$, while that from faulted and unfaulted areas scales
with $N^{2}$. Hence, for $N\sim\infty$, the contribution of dimers
can be neglected and the surface energy is simply the average between
energies of two $(111)\textrm{-}2\times2$ surfaces composed of adatoms:
with and without stacking faults, respectively.

The surface energy difference between $9\times9$ and $7\times7$
reconstructed Si$(111)$ surfaces was estimated from experimental
STM data in Ref.~\onlinecite{will94} as less than $4\,\mathrm{meV/1\times1\,cell}$
($0.3\,\mathrm{meV/\mathring{A}^{2}}$). This figure is somewhat lower
but comparable to the calculated energy differences of $0.42\,\mathrm{meV/\mathring{A}^{2}}$
(LDA) and $0.47\,\mathrm{meV/\mathring{A}^{2}}$ (GGA) as derived
from Tab.~\ref{tab1}. According to Tab.~\ref{tab1}, the $3\times3$
reconstructed Si$(111)$ surface has a significantly higher surface
energy than $5\times5$, $7\times7$ and $9\times9$ reconstructed
surfaces. The Si$(111)\textrm{-}5\times5$ surface has the lowest
surface energy, and that increases monotonically with increasing the
size of the unit cell. The surface energy of the infinitely large
DAS reconstruction is substantially (about $1.5\,\mathrm{meV/\mathring{A}^{2}}$)
higher than that of $3\times3$.

It is well known from experiments that $7\times7$ is the most frequently
observed surface reconstruction of Si$(111)$ at temperatures below
the $7\times7\leftrightarrow1\times1$ order-disorder structural transition.
Therefore, the data of Tab.~\ref{tab1} are in seeming disagreement
with the experimental results. However, considering the small energy
difference between $5\times5$ and $7\times7$ reconstructed surfaces,
it is important to inspect the relative stability of both surfaces
at finite temperatures, accounting for contributions of vibrational
and electronic excitations to the surface free energy.

\subsection{Contribution of vibrational and electronic excitations to the free
energies of Si$(111)-5\times5$ and Si$(111)-7\times7$ surfaces}

The vibrational free energy of $5\times5$ and $7\times7$ reconstructed
Si$(111)$ surfaces was obtained within the quasi-harmonic approximation,
according to the usual procedure employed for defects and surfaces
in semiconductors and metals \citep{san03,est04,mur15,kem19}. This
approach is applicable to silicon up to few hundred Kelvin \citep{est04,gom22},
and in particular to Si(111) DAS surfaces, which form unique and stable
minimum energy structures. For surfaces with structural degrees of
freedom, \emph{e.g.} Si$(100)$ \citep{sei04} and Si$(331)$ \citep{zha18}
one would have to consider configurational entropy and anharmonicity
\citep{kem19,for21}. For those cases, the potential energy surface
contains multiple minima separated by low energy barriers due to two-fold
buckling of small Si structures (forming double well potentials for
the case of buckled dimers on Si$(100)$).

The relative surface free energy can be calculated using Eq.~\ref{eq:delta_gamma},
where $\Delta f_{N\times N}=\Delta\gamma_{N\times N}+\Delta f_{N\times N,\textrm{vib}}$
now includes a vibrational contribution (in addition to the electronic
potential difference $\Delta\gamma_{N\times N}$ already reported
in the previous section). The quantity $\Delta f_{N\times N,\textrm{vib}}$
can be calculated by replacing $F_{N\times N}$ in Eq.~\ref{eq:delta_gamma}
by the vibrational contribution to the Helmholtz free energy of a
slab \citep{ful10},

\begin{equation}
F_{N\times N,\textrm{vib}}=\underset{k}{\sum}\frac{\hbar\omega_{k}}{2}+k_{\textrm{B}}T\underset{k}{\sum}\ln\left[1-\exp\left(-\frac{\hbar\omega_{k}}{k_{\textrm{B}}T}\right)\right],\label{eq:free_vib}
\end{equation}
and $\mu=\mu_{\textrm{vib}}$ is now the vibrational part of the Si
chemical potential (see below). The summation of Eq.~\ref{eq:free_vib}
runs over all vibrational mode frequencies, $\omega_{k}$, of the
slab (excludes 3 translational modes), and $k_{\textrm{B}}$ is the
Boltzmann constant. The first term on the right side of Eq.~\ref{eq:free_vib}
gives the ZPE. This is a $T$-independent quantity, which could in
principle be regarded as a potential term, and alternatively be considered
as a small correction to the $\gamma$ values of Tab.~\ref{tab1}.

The phonon frequencies were calculated using the frozen phonon scheme
as implemented in the \texttt{VIBRA} utility of the \texttt{SIESTA}
software package \citep{gar20}. Only zone-center phonons ($\mathbf{q}=\Gamma$)
were calculated. This is justified by the large lateral dimensions
of $5\times5$ and $7\times7$ reconstructions ($a_{5\times5}\approx19\,\mathrm{\mathring{A}}$,
$a_{7\times7}\approx27\,\mathrm{\mathring{A}}$), corresponding to
small BZ surface areas and rather weak phonon dispersion. For the
calculation of the dynamical matrix, we employed 4~BL-thick slabs
($\approx10\,\mathrm{\mathring{A}}$ thickness). All Si atoms of the
three top-most BLs were displaced by $0.02\,\mathrm{\mathring{A}}$
from their relaxed positions along all three Cartesian coordinates.
These correspond to a total of 150 and 298 dynamical atoms for the
$5\times5$ and $7\times7$ reconstructions, respectively. Thicker
slabs would require a computational power out of our reach. However,
as we will show below, the calculated vibrational free energy of a
64-atom supercell (with a (super)lattice spacing $a\approx11\,\mathrm{\mathring{A}}$),
is already converged and the corresponding vibrational entropy agrees
well with the measurements.

\begin{figure}
\includegraphics[clip,width=8.5cm]{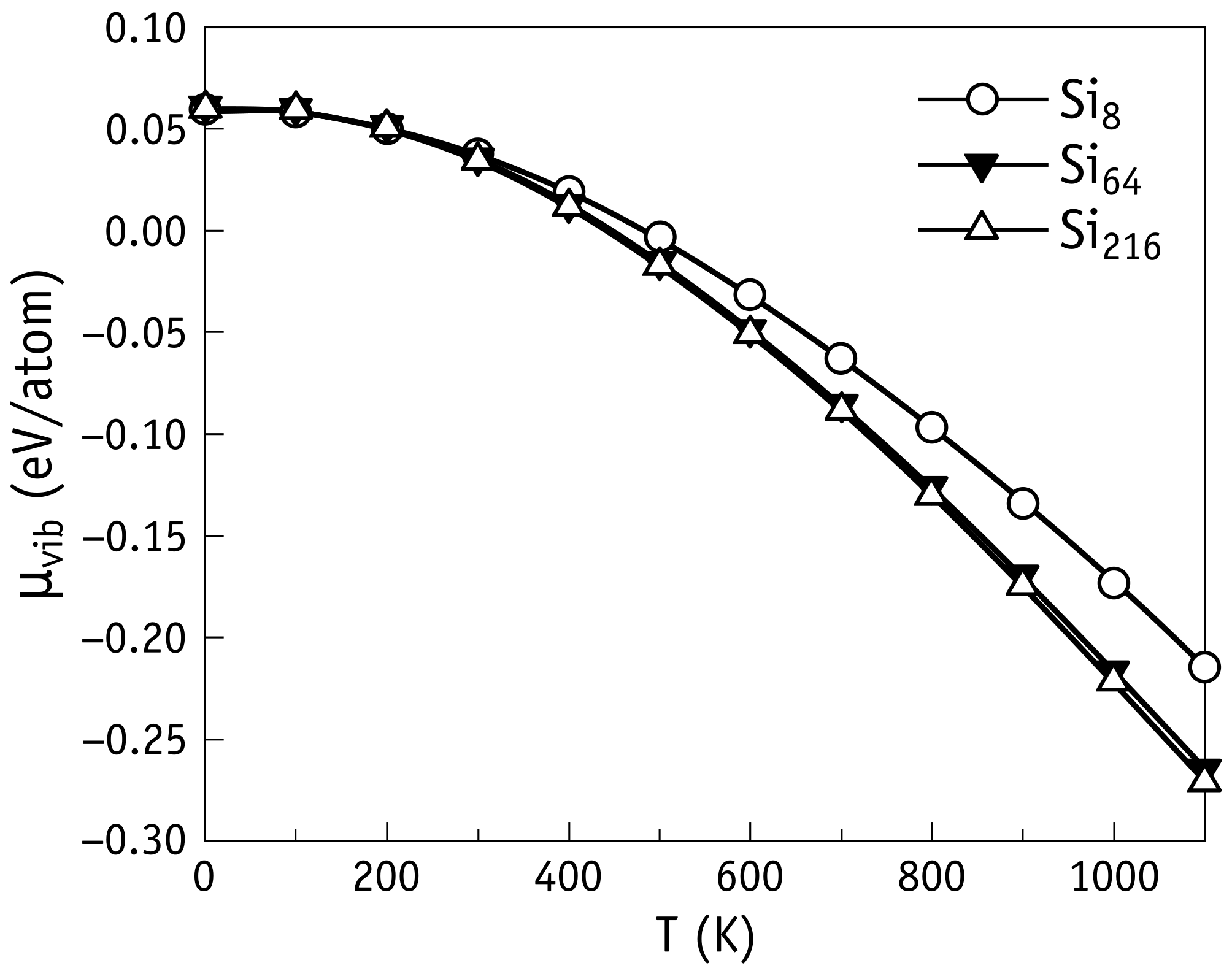}

\caption{\label{fig2} Calculated vibrational contribution to the Si chemical
potential $\mu_{\textrm{vib}}(T)$ using different supercell sizes
(see legend) as a function of temperature. The XC functional employed
was GGA PBEsol.}
\end{figure}

The vibrational contribution to the Si chemical potential, $\mu_{\textrm{vib}}$,
was calculated using Eq.~\ref{eq:free_vib} for a cubic bulk Si cell.
In order to test the convergence with respect to the cell size, we
performed calculations using $\mathrm{Si_{8}}$ ($a\approx5\,\mathrm{\mathring{A}}$),
$\mathrm{Si_{64}}$ ($a\approx11\,\mathrm{\mathring{A}}$) and $\mathrm{Si_{216}}$
($a\approx16\,\mathrm{\mathring{A}}$) bulk cubic cells. Results within
GGA PBEsol are shown in Fig.~\ref{fig2}; the LDA PZ results are
comparable. It is clear from this figure that the values for $\mathrm{Si_{64}}$
and $\mathrm{Si_{216}}$ clusters are very close: their difference
amounts to $\Delta\mu_{\textrm{vib}}=4\,\mathrm{meV/atom}$ at $T=800\,\mathrm{K}$,
which translates into an error bar of only $0.03\,\mathrm{meV/\mathring{A}^{2}}$
in the calculation of $\Delta f_{7\times7,\textrm{vib}}$ using Eq.~\ref{eq:delta_gamma}.
These findings are corroborated by the results of Estreicher et~al.
\citep{est04} and by convergence tests reported by Gomes \emph{et~al.}
\citep{gom22}, who found no significant improvement in the calculated
specific heat of Si when moving from a 64-atom cell to a 216-atom
cell.

\noindent 

We also calculated the ZPE and vibrational entropy per Si atom, $s_{\textrm{vib}}=-\left(\partial\mu_{\textrm{vib}}/\partial T\right)_{V}$,
for bulk Si. The results per molar unit are shown in Tab.~\ref{tab:2}
and they are close to both experimental data \citep{flu59,bar95}
and previous calculations \citep{san03}.

\noindent 
\begin{table}
\caption{\label{tab:2}Calculated (Calc.) and experimental (Exp.) values of
ZPE (kJ/mol) and vibrational entropy $s_{\textrm{vib}}\,(\textrm{J/K\ensuremath{\cdot}mol})$
of crystalline Si. The first two rows of calculated data correspond
to the present work.}

\begin{ruledtabular}
\begin{tabular}{lccc}
 & ZPE & $s_{\textrm{vib}}(300\,\textrm{K})$ & $s_{\textrm{vib}}(800\,\textrm{K})$\tabularnewline
\hline 
Calc.: LDA PZ, $\mathrm{Si_{216}}$ & 5.980 & 18.896 & 40.847\tabularnewline
Calc.: GGA PBEsol, $\mathrm{Si_{216}}$ & 5.871 & 19.327 & 41.350\tabularnewline
Calc.: LDA PZ, $\mathrm{Si_{64}}$ \citep{san03} & 6.201 & 18.395 & 40.308\tabularnewline
Exp. \citep{flu59,bar95} & 6.008 & 18.820 & 41.568\tabularnewline
\end{tabular}
\end{ruledtabular}

\end{table}

\begin{figure}
\includegraphics[clip,width=8.5cm]{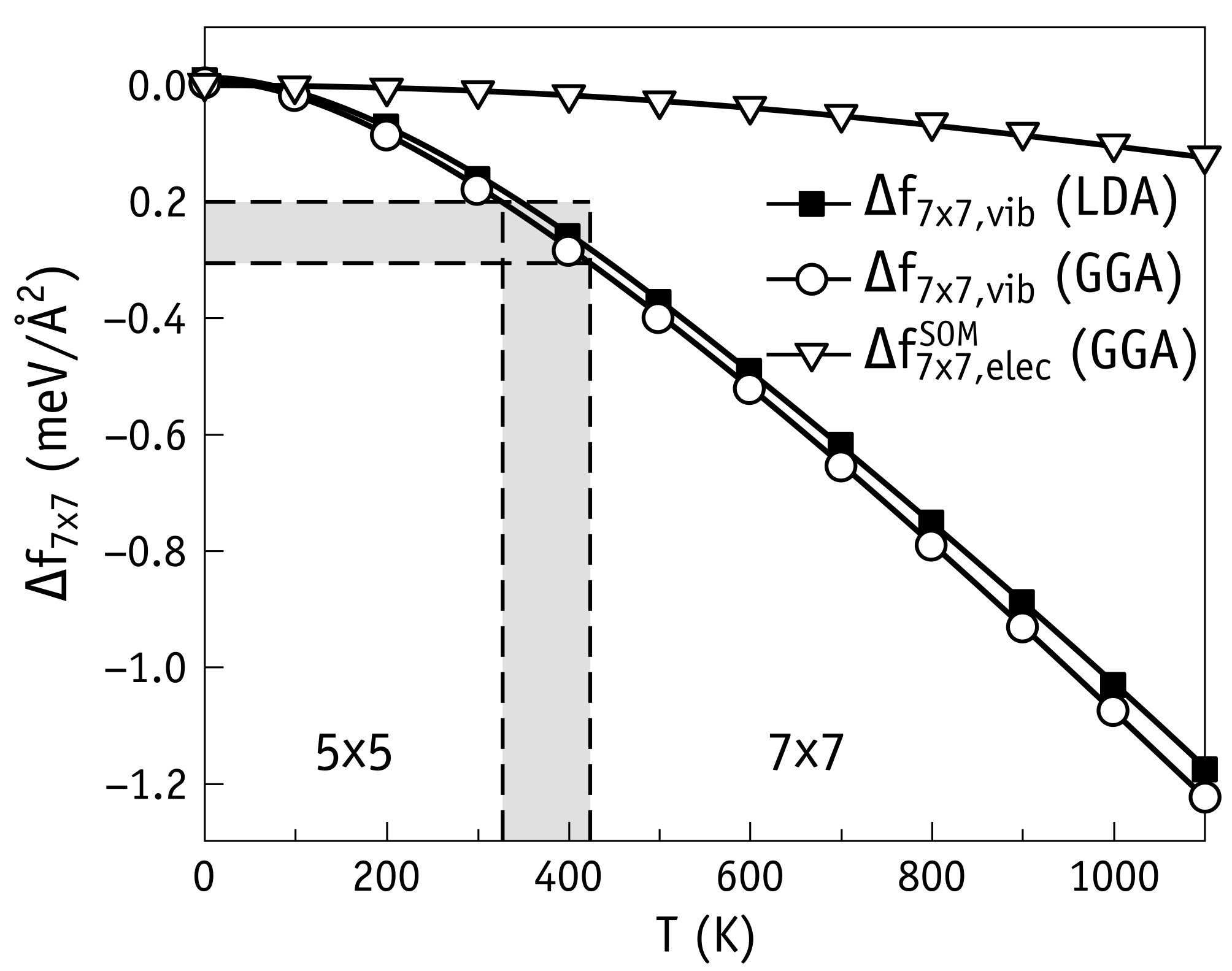}

\caption{\label{fig3} Calculated $\Delta f_{7\times7,\textrm{vib}}=f_{7\times7,\textrm{vib}}-f_{5\times5,\textrm{vib}}$
and $\Delta f_{7\times7,\textrm{elec}}^{\textrm{SOM}}=f_{7\times7,\textrm{elec}}^{\textrm{SOM}}-f_{5\times5,\textrm{elec}}^{\textrm{SOM}}$
as a function of temperature obtained within LDA PZ and GGA PBEsol.
The dashed lines highlight the ranges of temperatures and $\Delta f_{7\times7,\textrm{vib}}$
values where the phase transition $5\times5\leftrightarrow7\times7$
is expected to take place.}
\end{figure}

The calculated relative energy $\Delta f_{7\times7,\textrm{vib}}=f_{7\times7,\textrm{vib}}-f_{5\times5,\textrm{vib}}$
as a function of $T$ is shown in Fig.~\ref{fig3} for LDA PZ \citep{per81}
and GGA PBEsol \citep{per08} XC functionals; A $\mathrm{Si_{216}}$
supercell was used for the calculation of $\mu_{\textrm{vib}}$. At
a glance, LDA and GGA results are very similar and indicate a decrease
of $\Delta f_{7\times7,\textrm{vib}}$ (stabilization of the $7\times7$
reconstruction) with increasing the temperature. We can only explain
this feature with a faster increase of vibrational entropy with temperature
for the Si(111)-$7\times7$ surface (compared to that of Si(111)-$5\times5$).
The identification of specific vibrational modes and surface bonds
at the origin of this effect may not be simple. As shown by Murali
\emph{et~al.} \citep{mur15} for the case of defects in metals, the
change in vibrational entropy is not necessarily accounted for by
well-identifiable localized defect modes, but mostly by resonant modes
associated with vibrations in the strained regions surrounding the
defects. The lower their frequency, the more likely is for such resonances
to become thermally populated at a certain temperature, and therefore
the higher their contribution to the entropy change.

Our resultssuggest a relative softening of the Si(111)-$7\times7$
surface and subsurface bonds when compared to Si(111)-$5\times5$.
That would make the excited vibrational spectrum of the Si(111)-$7\times7$
more dense and accessible with temperature raising.

At this point, an obvious question is whether electronic entropy also
has a role to play? It has been shown that gap states deeper than
$\sim0.1$~eV from the band edges impact no more than few tens of
meV to the electronic free energy of formation, even at several hundred
Kelvin \citep{san03,est04}. Explicit calculations of the electronic
free energy of TiN semiconductor surface confirms that electronic
excitations are only significant above $T\approx600\textrm{-}700$\ K
\citep{for21}. Sommerfeld's approximation to the electronic free
energy of metals \citep{Ashcroft1976},

\begin{equation}
f_{\textrm{elec}}^{\textrm{SOM}}(T)=-\frac{\pi^{2}}{6}(k_{\textrm{B}}T)^{2}D(E_{\textrm{F}}),\label{eq:sommerfeld}
\end{equation}
provides us with an approximate and simple quantification of electronic
entropy of a metallic or semi-metallic surface. Accordingly, $f_{\textrm{elec}}^{\textrm{SOM}}$
is proportional to the density of electronic states at the Fermi level,
$D(E_{\textrm{F}})$. Even for a Si$(111)\textrm{-}7\times7$ reconstruction,
where the gap between highest occupied and lowest unoccupied surface
states can be as narrow as few tens of meV \citep{mod20}, $D(E_{\textrm{F}})$
is rather small, implying that the difference $\Delta D(E_{\textrm{F}})$
between $D(E_{\textrm{F}})$ of two close surfaces (such as $5\times5$
and $7\times7$ members of the DAS family), should be also small.
Sommerfeld's electronic free energy of Si(111)-$7\times7$ with respect
to the same quantity for Si(111)-$5\times5$, namely $\Delta f_{7\times7,\textrm{elec}}^{\textrm{SOM}}$,
was calculated using Eqs.~\ref{eq:delta_gamma} and \ref{eq:sommerfeld},
assuming a vanishing electronic entropy from the bulk Si chemical
potential ($\mu=\mu_{0}$). We applied a $0.1\,\mathrm{eV}$-wide
Gaussian broadening to the Kohn-Sham energy levels in the evaluation
of $D(E_{\textrm{F}})$. The results obtained within GGA PBEsol are
shown in Fig.~\ref{fig3} as open triangles. Like $\Delta f_{7\times7,\textrm{vib}}$,
the electronic surface free energy decreases with increasing the temperature
(also working toward a stabilization of the $7\times7$ reconstruction).
This is explained by approximately 8\% higher density of dangling
bonds on $7\times7$ reconstructed surface than on $5\times5$ surface,
and therefore by a higher density of states of the $7\times7$ reconstruction
within the Si gap. However, as anticipated, the calculated values
of $\Delta f_{7\times7,\textrm{elec}}^{\textrm{SOM}}$ are at least
one order of magnitude lower than $\Delta f_{7\times7,\textrm{vib}}$,
suggesting that electronic entropy can be safely neglected. Hence,
to a good approximation, the free energy difference between $7\times7$
and $5\times5$ reconstructed surfaces is simply given by $\Delta f_{7\times7}=\Delta\gamma_{7\times7}+\Delta f_{\textrm{vib},7\times7}$.

We reported already (see Tab.~\ref{tab1}) that the electronic surface
energy $\Delta\gamma_{7\times7}$ is positive and dominates the surface
free energy at low temperatures (favoring the $5\times5$ structure).
On the other hand, $\Delta f_{\textrm{vib},7\times7}$ (Fig.~\ref{fig3})
is increasingly negative with temperature (favoring the $7\times7$
structure), meaning that there is a phase transition temperature above
which the Si(111)-$7\times7$ becomes more stable. Considering that
$\Delta\gamma_{7\times7}\approx0.2\textrm{-}0.3\,\mathrm{meV/\mathring{A}^{2}}$
(Tab.~\ref{tab1}) the transition temperature is found at $T_{\textrm{trans}}\approx300\textrm{-}400\,\mathrm{K}$,
\emph{i.e.}, slightly above room temperature (Fig.~\ref{fig3}).

The usual way to prepare a clean Si(111) surface involves sample flashing
at $1250\,\mathrm{\lyxmathsym{\textdegree}C}$ for about $1\,\mathrm{min}$
followed by cooling to the desired temperature \citep{zha21}. Above
the $7\times7\leftrightarrow1\times1$ order-disorder phase transition
temperature at $T\approx1100\,\mathrm{K}$, the Si$(111)$ surface
contains a high density of mobile Si adatoms \citep{will94} and exhibits
$1\times1$ diffraction pattern reflecting the Si substrate below
the moving adatoms. The $7\times7$ reconstruction forms at around
the transition temperature and persists at lower temperatures. Our
results are in agreement with these well established observations,
namely, the Si(111)-$7\times7$ reconstruction is the most stable
across a wide temperature range up to $T=1100\,\mathrm{K}$.

However, the $7\times7$ structure on Si$(111)$ was also observed
below room temperature with no sign of structure change to $5\times5$
\citep{mod20}. The most likely explanation for this experimental
fact is the existence of a high kinetic barrier making the $5\times5\leftrightarrow7\times7$
transformation unlikely. The hindered diffusion of Si atoms on $5\times5$
and $7\times7$ reconstructed surfaces below room temperature provides
an example of an effective slowing factor. Other effects, including
a barrier for Si atom detachment may be invoked as well. Indeed, the
$5\times5$ and $7\times7$ reconstructions have different surface
atomic density \citep{will94}. Hence, the $5\times5\leftrightarrow7\times7$
structural transformation requires the removal (or addition) of excessive
(or deficient) Si surface atoms from Si$(111)$ terraces, for instance
by moving them to the always existing step edges on the sample surface.
Si surface diffusion on Si$(111)$-$7\times7$ at room temperature
is strongly hindered due to the high energy barrier ($E_{\textrm{b}}=1.14\,\mathrm{eV}$)
that separates two halves of the $7\times7$ unit cell \citep{sat00}.
This experimental figure is in agreement with DFT calculations showing
that there is an energy barrier of about $1\,\mathrm{eV}$ separating
faulted and unfaulted $7\times7$ half-cells for diffusing Si adatoms
\citep{cha03}. The height of energy barriers for adatom migration
depend on the local atomic arrangement of surface atoms within the
DAS structures. This is essentially identical for $5\times5$ and
$7\times7$ reconstructions, suggesting that the energy barriers on
$(111)\textrm{-}7\times7$ and -$5\times5$ surfaces should be similarly
high. Such a conclusion was also drawn from experimental STM data
obtained on Ge$(111)\textrm{-}7\times7$ and -$5\times5$ reconstructed
surfaces \citep{che04}. Thus, according to our calculations the $7\times7$
reconstruction is metastable below room temperature, being frozen
in upon cooling from high temperatures.

\noindent 

\section{Conclusions}

The relative thermodynamic stability of the $3\times3$, $5\times5$,
$7\times7$, $9\times9$ and infinitely large DAS reconstructions
on Si$(111)$ surface was investigated by means of first-principles
calculations within density functional theory. It was found that the
surface energy ordering (from lowest to highest) at $T=0\,\mathrm{K}$
is: $5\times5$, $7\times7$, $9\times9$, $3\times3$, infinitely
large. The vibrational entropy contribution to the free energy of
formation was evaluated for the $(111)\textrm{-}7\times7$ and $\textrm{-}5\times5$
surfaces in the temperature range $T=0\textrm{-}1100\,\mathrm{K}$.
The contribution of electronic entropy was shown to be negligible.
Accordingly, the $5\times5$ reconstruction is energetically more
favorable than $7\times7$ at low temperatures, with the $5\times5\leftrightarrow7\times7$
phase transition anticipated to occur at around room temperature.
The invariable observation of the $7\times7$ phase is explained by
the low mobility of Si atoms on the surface, effectively freezing
it in a metastable state during cooling. Our study shows that vibrational
entropy plays a crucial role in the stabilization of $7\times7$ reconstruction
at elevated temperatures, and reveals the metastable nature of this
structure below room temperature.
\begin{acknowledgments}
R. A. Z. would like to thank the Novosibirsk State University for
providing the computational resources. This work was supported by
the Russian Science Foundation (project no. 19-72-30023). J. C. thanks
the support of FCT in Portugal through Projects LA/P/0037/2020, UIDB/50025/2020,
UIDP/50025/2020
\end{acknowledgments}

\bibliographystyle{apsrev4-2}

\begin{thebibliography}{47}%
\makeatletter
\providecommand \@ifxundefined [1]{%
 \@ifx{#1\undefined}
}%
\providecommand \@ifnum [1]{%
 \ifnum #1\expandafter \@firstoftwo
 \else \expandafter \@secondoftwo
 \fi
}%
\providecommand \@ifx [1]{%
 \ifx #1\expandafter \@firstoftwo
 \else \expandafter \@secondoftwo
 \fi
}%
\providecommand \natexlab [1]{#1}%
\providecommand \enquote  [1]{``#1''}%
\providecommand \bibnamefont  [1]{#1}%
\providecommand \bibfnamefont [1]{#1}%
\providecommand \citenamefont [1]{#1}%
\providecommand \href@noop [0]{\@secondoftwo}%
\providecommand \href [0]{\begingroup \@sanitize@url \@href}%
\providecommand \@href[1]{\@@startlink{#1}\@@href}%
\providecommand \@@href[1]{\endgroup#1\@@endlink}%
\providecommand \@sanitize@url [0]{\catcode `\\12\catcode `\$12\catcode
  `\&12\catcode `\#12\catcode `\^12\catcode `\_12\catcode `\%12\relax}%
\providecommand \@@startlink[1]{}%
\providecommand \@@endlink[0]{}%
\providecommand \url  [0]{\begingroup\@sanitize@url \@url }%
\providecommand \@url [1]{\endgroup\@href {#1}{\urlprefix }}%
\providecommand \urlprefix  [0]{URL }%
\providecommand \Eprint [0]{\href }%
\providecommand \doibase [0]{https://doi.org/}%
\providecommand \selectlanguage [0]{\@gobble}%
\providecommand \bibinfo  [0]{\@secondoftwo}%
\providecommand \bibfield  [0]{\@secondoftwo}%
\providecommand \translation [1]{[#1]}%
\providecommand \BibitemOpen [0]{}%
\providecommand \bibitemStop [0]{}%
\providecommand \bibitemNoStop [0]{.\EOS\space}%
\providecommand \EOS [0]{\spacefactor3000\relax}%
\providecommand \BibitemShut  [1]{\csname bibitem#1\endcsname}%
\let\auto@bib@innerbib\@empty
\bibitem [{\citenamefont {D\c{a}browski}\ \emph {et~al.}(1994)\citenamefont
  {D\c{a}browski}, \citenamefont {M{\"u}ssig},\ and\ \citenamefont
  {Wolff}}]{dab94}%
  \BibitemOpen
  \bibfield  {author} {\bibinfo {author} {\bibfnamefont {J.}~\bibnamefont
  {D\c{a}browski}}, \bibinfo {author} {\bibfnamefont {H.-J.}\ \bibnamefont
  {M{\"u}ssig}},\ and\ \bibinfo {author} {\bibfnamefont {G.}~\bibnamefont
  {Wolff}},\ }\href {https://doi.org/10.1103/PhysRevLett.73.1660} {\bibfield
  {journal} {\bibinfo  {journal} {Phys. Rev. Lett.}\ }\textbf {\bibinfo
  {volume} {73}},\ \bibinfo {pages} {1660} (\bibinfo {year}
  {1994})}\BibitemShut {NoStop}%
\bibitem [{\citenamefont {Erwin}\ \emph {et~al.}(1996)\citenamefont {Erwin},
  \citenamefont {Baski},\ and\ \citenamefont {Whitman}}]{erw96}%
  \BibitemOpen
  \bibfield  {author} {\bibinfo {author} {\bibfnamefont {S.~C.}\ \bibnamefont
  {Erwin}}, \bibinfo {author} {\bibfnamefont {A.~A.}\ \bibnamefont {Baski}},\
  and\ \bibinfo {author} {\bibfnamefont {L.~J.}\ \bibnamefont {Whitman}},\
  }\href {https://doi.org/10.1103/PhysRevLett.77.687} {\bibfield  {journal}
  {\bibinfo  {journal} {Phys. Rev. Lett.}\ }\textbf {\bibinfo {volume} {77}},\
  \bibinfo {pages} {687} (\bibinfo {year} {1996})}\BibitemShut {NoStop}%
\bibitem [{\citenamefont {Bechstedt}\ \emph {et~al.}(2001)\citenamefont
  {Bechstedt}, \citenamefont {Stekolnikov}, \citenamefont {Furthm{\"u}ller},\
  and\ \citenamefont {K{\"a}ckell}}]{bech01}%
  \BibitemOpen
  \bibfield  {author} {\bibinfo {author} {\bibfnamefont {F.}~\bibnamefont
  {Bechstedt}}, \bibinfo {author} {\bibfnamefont {A.~A.}\ \bibnamefont
  {Stekolnikov}}, \bibinfo {author} {\bibfnamefont {J.}~\bibnamefont
  {Furthm{\"u}ller}},\ and\ \bibinfo {author} {\bibfnamefont {P.}~\bibnamefont
  {K{\"a}ckell}},\ }\href {https://doi.org/10.1103/PhysRevLett.87.016103}
  {\bibfield  {journal} {\bibinfo  {journal} {Phys. Rev. Lett.}\ }\textbf
  {\bibinfo {volume} {87}},\ \bibinfo {pages} {016103} (\bibinfo {year}
  {2001})}\BibitemShut {NoStop}%
\bibitem [{\citenamefont {Battaglia}\ \emph {et~al.}(2009)\citenamefont
  {Battaglia}, \citenamefont {Ga{\'a}l-Nagy}, \citenamefont {Monney},
  \citenamefont {Didiot}, \citenamefont {Schwier}, \citenamefont {Garnier},
  \citenamefont {Onida},\ and\ \citenamefont {Aebi}}]{bat09}%
  \BibitemOpen
  \bibfield  {author} {\bibinfo {author} {\bibfnamefont {C.}~\bibnamefont
  {Battaglia}}, \bibinfo {author} {\bibfnamefont {K.}~\bibnamefont
  {Ga{\'a}l-Nagy}}, \bibinfo {author} {\bibfnamefont {C.}~\bibnamefont
  {Monney}}, \bibinfo {author} {\bibfnamefont {C.}~\bibnamefont {Didiot}},
  \bibinfo {author} {\bibfnamefont {E.~F.}\ \bibnamefont {Schwier}}, \bibinfo
  {author} {\bibfnamefont {M.~G.}\ \bibnamefont {Garnier}}, \bibinfo {author}
  {\bibfnamefont {G.}~\bibnamefont {Onida}},\ and\ \bibinfo {author}
  {\bibfnamefont {P.}~\bibnamefont {Aebi}},\ }\href
  {https://doi.org/10.1103/PhysRevLett.102.066102} {\bibfield  {journal}
  {\bibinfo  {journal} {Phys. Rev. Lett.}\ }\textbf {\bibinfo {volume} {102}},\
  \bibinfo {pages} {066102} (\bibinfo {year} {2009})}\BibitemShut {NoStop}%
\bibitem [{\citenamefont {Kuzmin}\ \emph {et~al.}(2014)\citenamefont {Kuzmin},
  \citenamefont {Punkkinen}, \citenamefont {Laukkanen}, \citenamefont
  {L{\aa}ng}, \citenamefont {Dahl}, \citenamefont {Vitos},\ and\ \citenamefont
  {Kokko}}]{punkkinen14}%
  \BibitemOpen
  \bibfield  {author} {\bibinfo {author} {\bibfnamefont {M.}~\bibnamefont
  {Kuzmin}}, \bibinfo {author} {\bibfnamefont {M.~P.~J.}\ \bibnamefont
  {Punkkinen}}, \bibinfo {author} {\bibfnamefont {P.}~\bibnamefont
  {Laukkanen}}, \bibinfo {author} {\bibfnamefont {J.~J.~K.}\ \bibnamefont
  {L{\aa}ng}}, \bibinfo {author} {\bibfnamefont {J.}~\bibnamefont {Dahl}},
  \bibinfo {author} {\bibfnamefont {L.}~\bibnamefont {Vitos}},\ and\ \bibinfo
  {author} {\bibfnamefont {K.}~\bibnamefont {Kokko}},\ }\href
  {https://doi.org/10.1021/jp4082849} {\bibfield  {journal} {\bibinfo
  {journal} {J. Phys. Chem. C}\ }\textbf {\bibinfo {volume} {118}},\ \bibinfo
  {pages} {1894} (\bibinfo {year} {2014})}\BibitemShut {NoStop}%
\bibitem [{\citenamefont {Zhachuk}\ and\ \citenamefont {Teys}(2017)}]{zha17}%
  \BibitemOpen
  \bibfield  {author} {\bibinfo {author} {\bibfnamefont {R.~A.}\ \bibnamefont
  {Zhachuk}}\ and\ \bibinfo {author} {\bibfnamefont {S.~A.}\ \bibnamefont
  {Teys}},\ }\href {https://doi.org/10.1103/PhysRevB.95.041412} {\bibfield
  {journal} {\bibinfo  {journal} {Phys. Rev. B}\ }\textbf {\bibinfo {volume}
  {95}},\ \bibinfo {pages} {041412(R)} (\bibinfo {year} {2017})}\BibitemShut
  {NoStop}%
\bibitem [{\citenamefont {Zhachuk}\ and\ \citenamefont
  {Shklyaev}(2019)}]{zha19}%
  \BibitemOpen
  \bibfield  {author} {\bibinfo {author} {\bibfnamefont {R.~A.}\ \bibnamefont
  {Zhachuk}}\ and\ \bibinfo {author} {\bibfnamefont {A.~A.}\ \bibnamefont
  {Shklyaev}},\ }\href {https://doi.org/10.1016/j.apsusc.2019.07.144}
  {\bibfield  {journal} {\bibinfo  {journal} {Appl. Surf. Sci.}\ }\textbf
  {\bibinfo {volume} {494}},\ \bibinfo {pages} {46} (\bibinfo {year}
  {2019})}\BibitemShut {NoStop}%
\bibitem [{\citenamefont {Zhachuk}\ \emph {et~al.}(2020)\citenamefont
  {Zhachuk}, \citenamefont {Dolbak},\ and\ \citenamefont {Shklyaev}}]{zha20}%
  \BibitemOpen
  \bibfield  {author} {\bibinfo {author} {\bibfnamefont {R.~A.}\ \bibnamefont
  {Zhachuk}}, \bibinfo {author} {\bibfnamefont {A.~E.}\ \bibnamefont
  {Dolbak}},\ and\ \bibinfo {author} {\bibfnamefont {A.~A.}\ \bibnamefont
  {Shklyaev}},\ }\href {https://doi.org/10.1016/j.susc.2019.121549} {\bibfield
  {journal} {\bibinfo  {journal} {Surf. Sci.}\ }\textbf {\bibinfo {volume}
  {693}},\ \bibinfo {pages} {121549} (\bibinfo {year} {2020})}\BibitemShut
  {NoStop}%
\bibitem [{\citenamefont {Zhachuk}\ and\ \citenamefont
  {Coutinho}(2020)}]{zha20-2}%
  \BibitemOpen
  \bibfield  {author} {\bibinfo {author} {\bibfnamefont {R.~A.}\ \bibnamefont
  {Zhachuk}}\ and\ \bibinfo {author} {\bibfnamefont {J.}~\bibnamefont
  {Coutinho}},\ }\href {https://doi.org/10.1016/j.apsusc.2020.147507}
  {\bibfield  {journal} {\bibinfo  {journal} {Appl. Surf. Sci.}\ }\textbf
  {\bibinfo {volume} {533}},\ \bibinfo {pages} {147507} (\bibinfo {year}
  {2020})}\BibitemShut {NoStop}%
\bibitem [{\citenamefont {Zhachuk}\ \emph {et~al.}(2021)\citenamefont
  {Zhachuk}, \citenamefont {Rogilo}, \citenamefont {Petrov}, \citenamefont
  {Sheglov}, \citenamefont {Latyshev}, \citenamefont {Colonna},\ and\
  \citenamefont {Ronci}}]{zha21}%
  \BibitemOpen
  \bibfield  {author} {\bibinfo {author} {\bibfnamefont {R.~A.}\ \bibnamefont
  {Zhachuk}}, \bibinfo {author} {\bibfnamefont {D.~I.}\ \bibnamefont {Rogilo}},
  \bibinfo {author} {\bibfnamefont {A.~S.}\ \bibnamefont {Petrov}}, \bibinfo
  {author} {\bibfnamefont {D.~V.}\ \bibnamefont {Sheglov}}, \bibinfo {author}
  {\bibfnamefont {A.~V.}\ \bibnamefont {Latyshev}}, \bibinfo {author}
  {\bibfnamefont {S.}~\bibnamefont {Colonna}},\ and\ \bibinfo {author}
  {\bibfnamefont {F.}~\bibnamefont {Ronci}},\ }\href
  {https://doi.org/10.1103/PhysRevB.104.125437} {\bibfield  {journal} {\bibinfo
   {journal} {Phys. Rev. B}\ }\textbf {\bibinfo {volume} {104}},\ \bibinfo
  {pages} {125437} (\bibinfo {year} {2021})}\BibitemShut {NoStop}%
\bibitem [{\citenamefont {Stekolnikov}\ \emph {et~al.}(2002)\citenamefont
  {Stekolnikov}, \citenamefont {Furthm{\"u}ller},\ and\ \citenamefont
  {Bechstedt}}]{ste02}%
  \BibitemOpen
  \bibfield  {author} {\bibinfo {author} {\bibfnamefont {A.~A.}\ \bibnamefont
  {Stekolnikov}}, \bibinfo {author} {\bibfnamefont {J.}~\bibnamefont
  {Furthm{\"u}ller}},\ and\ \bibinfo {author} {\bibfnamefont {F.}~\bibnamefont
  {Bechstedt}},\ }\href {https://doi.org/10.1103/PhysRevB.65.115318} {\bibfield
   {journal} {\bibinfo  {journal} {Phys. Rev. B}\ }\textbf {\bibinfo {volume}
  {65}},\ \bibinfo {pages} {115318} (\bibinfo {year} {2002})}\BibitemShut
  {NoStop}%
\bibitem [{\citenamefont {Zhachuk}\ \emph {et~al.}(2013)\citenamefont
  {Zhachuk}, \citenamefont {Teys},\ and\ \citenamefont {Coutinho}}]{zha13}%
  \BibitemOpen
  \bibfield  {author} {\bibinfo {author} {\bibfnamefont {R.}~\bibnamefont
  {Zhachuk}}, \bibinfo {author} {\bibfnamefont {S.}~\bibnamefont {Teys}},\ and\
  \bibinfo {author} {\bibfnamefont {J.}~\bibnamefont {Coutinho}},\ }\href
  {https://doi.org/10.1063/1.4808356} {\bibfield  {journal} {\bibinfo
  {journal} {J. Chem. Phys.}\ }\textbf {\bibinfo {volume} {138}},\ \bibinfo
  {pages} {224702} (\bibinfo {year} {2013})}\BibitemShut {NoStop}%
\bibitem [{\citenamefont {Pandey}(1981)}]{pan81}%
  \BibitemOpen
  \bibfield  {author} {\bibinfo {author} {\bibfnamefont {K.~C.}\ \bibnamefont
  {Pandey}},\ }\href {https://doi.org/10.1103/PhysRevLett.47.1913} {\bibfield
  {journal} {\bibinfo  {journal} {Phys. Rev. Lett.}\ }\textbf {\bibinfo
  {volume} {47}},\ \bibinfo {pages} {1913} (\bibinfo {year}
  {1981})}\BibitemShut {NoStop}%
\bibitem [{\citenamefont {Zhachuk}\ \emph {et~al.}(2017)\citenamefont
  {Zhachuk}, \citenamefont {Coutinho}, \citenamefont {Dolbak}, \citenamefont
  {Cherepanov},\ and\ \citenamefont {Voigtl\"{a}nder}}]{zha17-2}%
  \BibitemOpen
  \bibfield  {author} {\bibinfo {author} {\bibfnamefont {R.}~\bibnamefont
  {Zhachuk}}, \bibinfo {author} {\bibfnamefont {J.}~\bibnamefont {Coutinho}},
  \bibinfo {author} {\bibfnamefont {A.}~\bibnamefont {Dolbak}}, \bibinfo
  {author} {\bibfnamefont {V.}~\bibnamefont {Cherepanov}},\ and\ \bibinfo
  {author} {\bibfnamefont {B.}~\bibnamefont {Voigtl\"{a}nder}},\ }\href
  {https://doi.org/10.1103/PhysRevB.96.085401} {\bibfield  {journal} {\bibinfo
  {journal} {Phys. Rev. B}\ }\textbf {\bibinfo {volume} {96}},\ \bibinfo
  {pages} {085401} (\bibinfo {year} {2017})}\BibitemShut {NoStop}%
\bibitem [{\citenamefont {Takayanagi}\ \emph {et~al.}(1985)\citenamefont
  {Takayanagi}, \citenamefont {Tanishiro}, \citenamefont {Takahashi},\ and\
  \citenamefont {Takahashi}}]{tak85}%
  \BibitemOpen
  \bibfield  {author} {\bibinfo {author} {\bibfnamefont {K.}~\bibnamefont
  {Takayanagi}}, \bibinfo {author} {\bibfnamefont {Y.}~\bibnamefont
  {Tanishiro}}, \bibinfo {author} {\bibfnamefont {S.}~\bibnamefont
  {Takahashi}},\ and\ \bibinfo {author} {\bibfnamefont {M.}~\bibnamefont
  {Takahashi}},\ }\href {https://doi.org/10.1016/0039-6028(85)90753-8}
  {\bibfield  {journal} {\bibinfo  {journal} {Surf. Sci.}\ }\textbf {\bibinfo
  {volume} {164}},\ \bibinfo {pages} {367} (\bibinfo {year}
  {1985})}\BibitemShut {NoStop}%
\bibitem [{\citenamefont {Seino}\ \emph {et~al.}(2004)\citenamefont {Seino},
  \citenamefont {Schmidt},\ and\ \citenamefont {Bechstedt}}]{sei04}%
  \BibitemOpen
  \bibfield  {author} {\bibinfo {author} {\bibfnamefont {K.}~\bibnamefont
  {Seino}}, \bibinfo {author} {\bibfnamefont {W.~G.}\ \bibnamefont {Schmidt}},\
  and\ \bibinfo {author} {\bibfnamefont {F.}~\bibnamefont {Bechstedt}},\ }\href
  {https://doi.org/10.1103/PhysRevLett.93.036101} {\bibfield  {journal}
  {\bibinfo  {journal} {Phys. Rev. Lett.}\ }\textbf {\bibinfo {volume} {93}},\
  \bibinfo {pages} {036101} (\bibinfo {year} {2004})}\BibitemShut {NoStop}%
\bibitem [{\citenamefont {\u{S}tich}\ \emph {et~al.}(1992)\citenamefont
  {\u{S}tich}, \citenamefont {Payne}, \citenamefont {King-Smith}, \citenamefont
  {Lin},\ and\ \citenamefont {Clarke}}]{sti92}%
  \BibitemOpen
  \bibfield  {author} {\bibinfo {author} {\bibfnamefont {I.}~\bibnamefont
  {\u{S}tich}}, \bibinfo {author} {\bibfnamefont {M.~C.}\ \bibnamefont
  {Payne}}, \bibinfo {author} {\bibfnamefont {R.~D.}\ \bibnamefont
  {King-Smith}}, \bibinfo {author} {\bibfnamefont {J.-S.}\ \bibnamefont
  {Lin}},\ and\ \bibinfo {author} {\bibfnamefont {L.~J.}\ \bibnamefont
  {Clarke}},\ }\href {https://doi.org/10.1103/PhysRevLett.68.1351} {\bibfield
  {journal} {\bibinfo  {journal} {Phys. Rev. Lett.}\ }\textbf {\bibinfo
  {volume} {68}},\ \bibinfo {pages} {1351} (\bibinfo {year}
  {1992})}\BibitemShut {NoStop}%
\bibitem [{\citenamefont {Ishizaka}\ \emph {et~al.}(1991)\citenamefont
  {Ishizaka}, \citenamefont {Doi},\ and\ \citenamefont {Ichikawa}}]{ish91}%
  \BibitemOpen
  \bibfield  {author} {\bibinfo {author} {\bibfnamefont {A.}~\bibnamefont
  {Ishizaka}}, \bibinfo {author} {\bibfnamefont {T.}~\bibnamefont {Doi}},\ and\
  \bibinfo {author} {\bibfnamefont {M.}~\bibnamefont {Ichikawa}},\ }\href
  {https://doi.org/10.1063/1.104471} {\bibfield  {journal} {\bibinfo  {journal}
  {Appl. Phys. Lett.}\ }\textbf {\bibinfo {volume} {58}},\ \bibinfo {pages}
  {902} (\bibinfo {year} {1991})}\BibitemShut {NoStop}%
\bibitem [{\citenamefont {Needels}(1993)}]{nee93}%
  \BibitemOpen
  \bibfield  {author} {\bibinfo {author} {\bibfnamefont {M.}~\bibnamefont
  {Needels}},\ }\href {https://doi.org/10.1103/PhysRevLett.71.3612} {\bibfield
  {journal} {\bibinfo  {journal} {Phys. Rev. Lett.}\ }\textbf {\bibinfo
  {volume} {71}},\ \bibinfo {pages} {3612} (\bibinfo {year}
  {1993})}\BibitemShut {NoStop}%
\bibitem [{\citenamefont {Solares}\ \emph {et~al.}(2005)\citenamefont
  {Solares}, \citenamefont {Dasgupta}, \citenamefont {Schultz}, \citenamefont
  {Kim}, \citenamefont {Musgrave},\ and\ \citenamefont {Goddard}}]{sol05}%
  \BibitemOpen
  \bibfield  {author} {\bibinfo {author} {\bibfnamefont {S.~D.}\ \bibnamefont
  {Solares}}, \bibinfo {author} {\bibfnamefont {S.}~\bibnamefont {Dasgupta}},
  \bibinfo {author} {\bibfnamefont {P.~A.}\ \bibnamefont {Schultz}}, \bibinfo
  {author} {\bibfnamefont {Y.-H.}\ \bibnamefont {Kim}}, \bibinfo {author}
  {\bibfnamefont {C.~B.}\ \bibnamefont {Musgrave}},\ and\ \bibinfo {author}
  {\bibfnamefont {W.~A.}\ \bibnamefont {Goddard}},\ }\href
  {https://doi.org/10.1021/la052029s} {\bibfield  {journal} {\bibinfo
  {journal} {Langmuir}\ }\textbf {\bibinfo {volume} {21}},\ \bibinfo {pages}
  {12404} (\bibinfo {year} {2005})}\BibitemShut {NoStop}%
\bibitem [{\citenamefont {Troullier}\ and\ \citenamefont
  {Martins}(1991)}]{tro91}%
  \BibitemOpen
  \bibfield  {author} {\bibinfo {author} {\bibfnamefont {N.}~\bibnamefont
  {Troullier}}\ and\ \bibinfo {author} {\bibfnamefont {J.~L.}\ \bibnamefont
  {Martins}},\ }\href {https://doi.org/10.1103/PhysRevB.43.1993} {\bibfield
  {journal} {\bibinfo  {journal} {Phys. Rev. B}\ }\textbf {\bibinfo {volume}
  {43}},\ \bibinfo {pages} {1993} (\bibinfo {year} {1991})}\BibitemShut
  {NoStop}%
\bibitem [{\citenamefont {Soler}\ \emph {et~al.}(2002)\citenamefont {Soler},
  \citenamefont {Artacho}, \citenamefont {Gale}, \citenamefont {Garc{\'\i}a},
  \citenamefont {Junquera}, \citenamefont {Ordej{\'o}n},\ and\ \citenamefont
  {S{\'a}nchez-Portal}}]{sol02}%
  \BibitemOpen
  \bibfield  {author} {\bibinfo {author} {\bibfnamefont {J.~M.}\ \bibnamefont
  {Soler}}, \bibinfo {author} {\bibfnamefont {E.}~\bibnamefont {Artacho}},
  \bibinfo {author} {\bibfnamefont {J.~D.}\ \bibnamefont {Gale}}, \bibinfo
  {author} {\bibfnamefont {A.}~\bibnamefont {Garc{\'\i}a}}, \bibinfo {author}
  {\bibfnamefont {J.}~\bibnamefont {Junquera}}, \bibinfo {author}
  {\bibfnamefont {P.}~\bibnamefont {Ordej{\'o}n}},\ and\ \bibinfo {author}
  {\bibfnamefont {D.}~\bibnamefont {S{\'a}nchez-Portal}},\ }\href
  {https://doi.org/10.1088/0953-8984/14/11/302} {\bibfield  {journal} {\bibinfo
   {journal} {J. Phys.: Condens. Matter}\ }\textbf {\bibinfo {volume} {14}},\
  \bibinfo {pages} {2745} (\bibinfo {year} {2002})}\BibitemShut {NoStop}%
\bibitem [{\citenamefont {Garc\'{i}a}\ \emph {et~al.}(2020)\citenamefont
  {Garc\'{i}a}, \citenamefont {Papior}, \citenamefont {Akhtar}, \citenamefont
  {Artacho}, \citenamefont {Blum}, \citenamefont {Bosoni}, \citenamefont
  {Brandimarte}, \citenamefont {Brandbyge}, \citenamefont {Cerd\'{a}},
  \citenamefont {Corsetti}, \citenamefont {Cuadrado}, \citenamefont {Dikan},
  \citenamefont {Ferrer}, \citenamefont {Gale}, \citenamefont
  {Garc\'{i}a-Fern\'{a}ndez}, \citenamefont {Garc\'{i}a-Su\'{a}rez},
  \citenamefont {Garc\'{i}a}, \citenamefont {Huhs}, \citenamefont {Illera},
  \citenamefont {Koryt\'{a}r}, \citenamefont {Koval}, \citenamefont {Lebedeva},
  \citenamefont {Lin}, \citenamefont {L\'{o}pez-Tarifa}, \citenamefont {Mayo},
  \citenamefont {Mohr}, \citenamefont {Ordej\'{o}n}, \citenamefont {Postnikov},
  \citenamefont {Pouillon}, \citenamefont {Pruneda}, \citenamefont {Robles},
  \citenamefont {S\'{a}nchez-Portal}, \citenamefont {Soler}, \citenamefont
  {Ullah}, \citenamefont {zhe Yu},\ and\ \citenamefont {Junquera}}]{gar20}%
  \BibitemOpen
  \bibfield  {author} {\bibinfo {author} {\bibfnamefont {A.}~\bibnamefont
  {Garc\'{i}a}}, \bibinfo {author} {\bibfnamefont {N.}~\bibnamefont {Papior}},
  \bibinfo {author} {\bibfnamefont {A.}~\bibnamefont {Akhtar}}, \bibinfo
  {author} {\bibfnamefont {E.}~\bibnamefont {Artacho}}, \bibinfo {author}
  {\bibfnamefont {V.}~\bibnamefont {Blum}}, \bibinfo {author} {\bibfnamefont
  {E.}~\bibnamefont {Bosoni}}, \bibinfo {author} {\bibfnamefont
  {P.}~\bibnamefont {Brandimarte}}, \bibinfo {author} {\bibfnamefont
  {M.}~\bibnamefont {Brandbyge}}, \bibinfo {author} {\bibfnamefont {J.~I.}\
  \bibnamefont {Cerd\'{a}}}, \bibinfo {author} {\bibfnamefont {F.}~\bibnamefont
  {Corsetti}}, \bibinfo {author} {\bibfnamefont {R.}~\bibnamefont {Cuadrado}},
  \bibinfo {author} {\bibfnamefont {V.}~\bibnamefont {Dikan}}, \bibinfo
  {author} {\bibfnamefont {J.}~\bibnamefont {Ferrer}}, \bibinfo {author}
  {\bibfnamefont {J.}~\bibnamefont {Gale}}, \bibinfo {author} {\bibfnamefont
  {P.}~\bibnamefont {Garc\'{i}a-Fern\'{a}ndez}}, \bibinfo {author}
  {\bibfnamefont {V.~M.}\ \bibnamefont {Garc\'{i}a-Su\'{a}rez}}, \bibinfo
  {author} {\bibfnamefont {S.}~\bibnamefont {Garc\'{i}a}}, \bibinfo {author}
  {\bibfnamefont {G.}~\bibnamefont {Huhs}}, \bibinfo {author} {\bibfnamefont
  {S.}~\bibnamefont {Illera}}, \bibinfo {author} {\bibfnamefont
  {R.}~\bibnamefont {Koryt\'{a}r}}, \bibinfo {author} {\bibfnamefont
  {P.}~\bibnamefont {Koval}}, \bibinfo {author} {\bibfnamefont
  {I.}~\bibnamefont {Lebedeva}}, \bibinfo {author} {\bibfnamefont
  {L.}~\bibnamefont {Lin}}, \bibinfo {author} {\bibfnamefont {P.}~\bibnamefont
  {L\'{o}pez-Tarifa}}, \bibinfo {author} {\bibfnamefont {S.~G.}\ \bibnamefont
  {Mayo}}, \bibinfo {author} {\bibfnamefont {S.}~\bibnamefont {Mohr}}, \bibinfo
  {author} {\bibfnamefont {P.}~\bibnamefont {Ordej\'{o}n}}, \bibinfo {author}
  {\bibfnamefont {A.}~\bibnamefont {Postnikov}}, \bibinfo {author}
  {\bibfnamefont {Y.}~\bibnamefont {Pouillon}}, \bibinfo {author}
  {\bibfnamefont {M.}~\bibnamefont {Pruneda}}, \bibinfo {author} {\bibfnamefont
  {R.}~\bibnamefont {Robles}}, \bibinfo {author} {\bibfnamefont
  {D.}~\bibnamefont {S\'{a}nchez-Portal}}, \bibinfo {author} {\bibfnamefont
  {J.~M.}\ \bibnamefont {Soler}}, \bibinfo {author} {\bibfnamefont
  {R.}~\bibnamefont {Ullah}}, \bibinfo {author} {\bibfnamefont {V.~W.}\
  \bibnamefont {zhe Yu}},\ and\ \bibinfo {author} {\bibfnamefont
  {J.}~\bibnamefont {Junquera}},\ }\href {https://doi.org/10.1063/5.0005077}
  {\bibfield  {journal} {\bibinfo  {journal} {J. Chem. Phys.}\ }\textbf
  {\bibinfo {volume} {152}},\ \bibinfo {pages} {204108} (\bibinfo {year}
  {2020})}\BibitemShut {NoStop}%
\bibitem [{\citenamefont {Perdew}\ and\ \citenamefont {Zunger}(1981)}]{per81}%
  \BibitemOpen
  \bibfield  {author} {\bibinfo {author} {\bibfnamefont {J.~P.}\ \bibnamefont
  {Perdew}}\ and\ \bibinfo {author} {\bibfnamefont {A.}~\bibnamefont
  {Zunger}},\ }\href {https://doi.org/10.1103/PhysRevB.23.5048} {\bibfield
  {journal} {\bibinfo  {journal} {Phys. Rev. B}\ }\textbf {\bibinfo {volume}
  {23}},\ \bibinfo {pages} {5048} (\bibinfo {year} {1981})}\BibitemShut
  {NoStop}%
\bibitem [{\citenamefont {Perdew}\ \emph {et~al.}(2008)\citenamefont {Perdew},
  \citenamefont {Ruzsinszky}, \citenamefont {Csonka}, \citenamefont {Vydrov},
  \citenamefont {Scuseria}, \citenamefont {Constantin}, \citenamefont {Zhou},\
  and\ \citenamefont {Burke}}]{per08}%
  \BibitemOpen
  \bibfield  {author} {\bibinfo {author} {\bibfnamefont {J.~P.}\ \bibnamefont
  {Perdew}}, \bibinfo {author} {\bibfnamefont {A.}~\bibnamefont {Ruzsinszky}},
  \bibinfo {author} {\bibfnamefont {G.~I.}\ \bibnamefont {Csonka}}, \bibinfo
  {author} {\bibfnamefont {O.~A.}\ \bibnamefont {Vydrov}}, \bibinfo {author}
  {\bibfnamefont {G.~E.}\ \bibnamefont {Scuseria}}, \bibinfo {author}
  {\bibfnamefont {L.~A.}\ \bibnamefont {Constantin}}, \bibinfo {author}
  {\bibfnamefont {X.}~\bibnamefont {Zhou}},\ and\ \bibinfo {author}
  {\bibfnamefont {K.}~\bibnamefont {Burke}},\ }\href
  {https://doi.org/10.1103/PhysRevLett.100.136406} {\bibfield  {journal}
  {\bibinfo  {journal} {Phys. Rev. Lett.}\ }\textbf {\bibinfo {volume} {100}},\
  \bibinfo {pages} {136406} (\bibinfo {year} {2008})}\BibitemShut {NoStop}%
\bibitem [{\citenamefont {Monkhorst}\ and\ \citenamefont {Pack}(1976)}]{mon76}%
  \BibitemOpen
  \bibfield  {author} {\bibinfo {author} {\bibfnamefont {H.~J.}\ \bibnamefont
  {Monkhorst}}\ and\ \bibinfo {author} {\bibfnamefont {J.~D.}\ \bibnamefont
  {Pack}},\ }\href {https://doi.org/10.1103/PhysRevB.13.5188} {\bibfield
  {journal} {\bibinfo  {journal} {Phys. Rev. B}\ }\textbf {\bibinfo {volume}
  {13}},\ \bibinfo {pages} {5188} (\bibinfo {year} {1976})}\BibitemShut
  {NoStop}%
\bibitem [{\citenamefont {Sanati}\ and\ \citenamefont
  {Estreicher}(2003)}]{san03}%
  \BibitemOpen
  \bibfield  {author} {\bibinfo {author} {\bibfnamefont {M.}~\bibnamefont
  {Sanati}}\ and\ \bibinfo {author} {\bibfnamefont {S.~K.}\ \bibnamefont
  {Estreicher}},\ }\href {https://doi.org/10.1016/j.ssc.2003.08.005} {\bibfield
   {journal} {\bibinfo  {journal} {Solid State Commun.}\ }\textbf {\bibinfo
  {volume} {128}},\ \bibinfo {pages} {181} (\bibinfo {year}
  {2003})}\BibitemShut {NoStop}%
\bibitem [{\citenamefont {Estreicher}\ \emph {et~al.}(2004)\citenamefont
  {Estreicher}, \citenamefont {Sanati}, \citenamefont {West},\ and\
  \citenamefont {Ruymgaart}}]{est04}%
  \BibitemOpen
  \bibfield  {author} {\bibinfo {author} {\bibfnamefont {S.~K.}\ \bibnamefont
  {Estreicher}}, \bibinfo {author} {\bibfnamefont {M.}~\bibnamefont {Sanati}},
  \bibinfo {author} {\bibfnamefont {D.}~\bibnamefont {West}},\ and\ \bibinfo
  {author} {\bibfnamefont {F.}~\bibnamefont {Ruymgaart}},\ }\href
  {https://doi.org/10.1103/PhysRevB.70.125209} {\bibfield  {journal} {\bibinfo
  {journal} {Phys. Rev. B}\ }\textbf {\bibinfo {volume} {70}},\ \bibinfo
  {pages} {125209} (\bibinfo {year} {2004})}\BibitemShut {NoStop}%
\bibitem [{\citenamefont {Kresse}\ and\ \citenamefont
  {Furthm\"{u}ller}(1996{\natexlab{a}})}]{kre96}%
  \BibitemOpen
  \bibfield  {author} {\bibinfo {author} {\bibfnamefont {G.}~\bibnamefont
  {Kresse}}\ and\ \bibinfo {author} {\bibfnamefont {J.}~\bibnamefont
  {Furthm\"{u}ller}},\ }\href {https://doi.org/10.1103/PhysRevB.54.11169}
  {\bibfield  {journal} {\bibinfo  {journal} {Phys. Rev. B}\ }\textbf {\bibinfo
  {volume} {54}},\ \bibinfo {pages} {11169} (\bibinfo {year}
  {1996}{\natexlab{a}})}\BibitemShut {NoStop}%
\bibitem [{\citenamefont {Kresse}\ and\ \citenamefont
  {Furthm\"{u}ller}(1996{\natexlab{b}})}]{kre96-2}%
  \BibitemOpen
  \bibfield  {author} {\bibinfo {author} {\bibfnamefont {G.}~\bibnamefont
  {Kresse}}\ and\ \bibinfo {author} {\bibfnamefont {J.}~\bibnamefont
  {Furthm\"{u}ller}},\ }\href {https://doi.org/10.1016/0927-0256(96)00008-0}
  {\bibfield  {journal} {\bibinfo  {journal} {Comp. Mater. Sci.}\ }\textbf
  {\bibinfo {volume} {6}},\ \bibinfo {pages} {15} (\bibinfo {year}
  {1996}{\natexlab{b}})}\BibitemShut {NoStop}%
\bibitem [{\citenamefont {Perdew}\ \emph {et~al.}(1996)\citenamefont {Perdew},
  \citenamefont {Burke},\ and\ \citenamefont {Ernzerhof}}]{per96}%
  \BibitemOpen
  \bibfield  {author} {\bibinfo {author} {\bibfnamefont {J.~P.}\ \bibnamefont
  {Perdew}}, \bibinfo {author} {\bibfnamefont {K.}~\bibnamefont {Burke}},\ and\
  \bibinfo {author} {\bibfnamefont {M.}~\bibnamefont {Ernzerhof}},\ }\href
  {https://doi.org/10.1103/PhysRevLett.77.3865} {\bibfield  {journal} {\bibinfo
   {journal} {Phys. Rev. Lett.}\ }\textbf {\bibinfo {volume} {77}},\ \bibinfo
  {pages} {3865} (\bibinfo {year} {1996})}\BibitemShut {NoStop}%
\bibitem [{\citenamefont {Grimmea}\ \emph {et~al.}(2010)\citenamefont
  {Grimmea}, \citenamefont {Antony}, \citenamefont {Ehrlich},\ and\
  \citenamefont {Krieg}}]{gri10}%
  \BibitemOpen
  \bibfield  {author} {\bibinfo {author} {\bibfnamefont {S.}~\bibnamefont
  {Grimmea}}, \bibinfo {author} {\bibfnamefont {J.}~\bibnamefont {Antony}},
  \bibinfo {author} {\bibfnamefont {S.}~\bibnamefont {Ehrlich}},\ and\ \bibinfo
  {author} {\bibfnamefont {H.}~\bibnamefont {Krieg}},\ }\href
  {https://doi.org/10.1063/1.3382344} {\bibfield  {journal} {\bibinfo
  {journal} {J. Chem. Phys.}\ }\textbf {\bibinfo {volume} {132}},\ \bibinfo
  {pages} {154104} (\bibinfo {year} {2010})}\BibitemShut {NoStop}%
\bibitem [{\citenamefont {Grimme}\ \emph {et~al.}(2011)\citenamefont {Grimme},
  \citenamefont {Ehrlich},\ and\ \citenamefont {Goerigk}}]{gri11}%
  \BibitemOpen
  \bibfield  {author} {\bibinfo {author} {\bibfnamefont {S.}~\bibnamefont
  {Grimme}}, \bibinfo {author} {\bibfnamefont {S.}~\bibnamefont {Ehrlich}},\
  and\ \bibinfo {author} {\bibfnamefont {L.}~\bibnamefont {Goerigk}},\ }\href
  {https://doi.org/10.1002/jcc.21759} {\bibfield  {journal} {\bibinfo
  {journal} {J. Comp. Chem.}\ }\textbf {\bibinfo {volume} {32}},\ \bibinfo
  {pages} {1456} (\bibinfo {year} {2011})}\BibitemShut {NoStop}%
\bibitem [{\citenamefont {Yang}\ and\ \citenamefont {Williams}(1994)}]{will94}%
  \BibitemOpen
  \bibfield  {author} {\bibinfo {author} {\bibfnamefont {Y.-N.}\ \bibnamefont
  {Yang}}\ and\ \bibinfo {author} {\bibfnamefont {E.~D.}\ \bibnamefont
  {Williams}},\ }\href {https://doi.org/10.1103/PhysRevLett.72.1862} {\bibfield
   {journal} {\bibinfo  {journal} {Phys. Rev. Lett.}\ }\textbf {\bibinfo
  {volume} {72}},\ \bibinfo {pages} {1862} (\bibinfo {year}
  {1994})}\BibitemShut {NoStop}%
\bibitem [{\citenamefont {Murali}\ \emph {et~al.}(2015)\citenamefont {Murali},
  \citenamefont {Posselt},\ and\ \citenamefont {Schiwarth}}]{mur15}%
  \BibitemOpen
  \bibfield  {author} {\bibinfo {author} {\bibfnamefont {D.}~\bibnamefont
  {Murali}}, \bibinfo {author} {\bibfnamefont {M.}~\bibnamefont {Posselt}},\
  and\ \bibinfo {author} {\bibfnamefont {M.}~\bibnamefont {Schiwarth}},\ }\href
  {https://doi.org/10.1103/physrevb.92.064103} {\bibfield  {journal} {\bibinfo
  {journal} {Physical Review B}\ }\textbf {\bibinfo {volume} {92}},\ \bibinfo
  {pages} {064103} (\bibinfo {year} {2015})}\BibitemShut {NoStop}%
\bibitem [{\citenamefont {Kempisty}\ and\ \citenamefont
  {Kangawa}(2019)}]{kem19}%
  \BibitemOpen
  \bibfield  {author} {\bibinfo {author} {\bibfnamefont {P.}~\bibnamefont
  {Kempisty}}\ and\ \bibinfo {author} {\bibfnamefont {Y.}~\bibnamefont
  {Kangawa}},\ }\href {https://doi.org/10.1103/physrevb.100.085304} {\bibfield
  {journal} {\bibinfo  {journal} {Physical Review B}\ }\textbf {\bibinfo
  {volume} {100}},\ \bibinfo {pages} {085304} (\bibinfo {year}
  {2019})}\BibitemShut {NoStop}%
\bibitem [{\citenamefont {Gomes}\ \emph {et~al.}(2022)\citenamefont {Gomes},
  \citenamefont {Markevich}, \citenamefont {Peaker},\ and\ \citenamefont
  {Coutinho}}]{gom22}%
  \BibitemOpen
  \bibfield  {author} {\bibinfo {author} {\bibfnamefont {D.}~\bibnamefont
  {Gomes}}, \bibinfo {author} {\bibfnamefont {V.~P.}\ \bibnamefont
  {Markevich}}, \bibinfo {author} {\bibfnamefont {A.~R.}\ \bibnamefont
  {Peaker}},\ and\ \bibinfo {author} {\bibfnamefont {J.}~\bibnamefont
  {Coutinho}},\ }\href {https://doi.org/10.1002/pssb.202100670} {\bibfield
  {journal} {\bibinfo  {journal} {Physica Status Solidi B}\ ,\ \bibinfo {pages}
  {2100670}} (\bibinfo {year} {2022})}\BibitemShut {NoStop}%
\bibitem [{\citenamefont {Zhachuk}\ \emph {et~al.}(2018)\citenamefont
  {Zhachuk}, \citenamefont {Coutinho},\ and\ \citenamefont
  {Palot\'{a}s}}]{zha18}%
  \BibitemOpen
  \bibfield  {author} {\bibinfo {author} {\bibfnamefont {R.}~\bibnamefont
  {Zhachuk}}, \bibinfo {author} {\bibfnamefont {J.}~\bibnamefont {Coutinho}},\
  and\ \bibinfo {author} {\bibfnamefont {K.}~\bibnamefont {Palot\'{a}s}},\
  }\href {https://doi.org/doi.org/10.1063/1.5048064} {\bibfield  {journal}
  {\bibinfo  {journal} {J. Chem. Phys.}\ }\textbf {\bibinfo {volume} {149}},\
  \bibinfo {pages} {204702} (\bibinfo {year} {2018})}\BibitemShut {NoStop}%
\bibitem [{\citenamefont {Forslund}\ \emph {et~al.}(2021)\citenamefont
  {Forslund}, \citenamefont {Zhang}, \citenamefont {Grabowski}, \citenamefont
  {Shapeev},\ and\ \citenamefont {Ruban}}]{for21}%
  \BibitemOpen
  \bibfield  {author} {\bibinfo {author} {\bibfnamefont {A.}~\bibnamefont
  {Forslund}}, \bibinfo {author} {\bibfnamefont {X.}~\bibnamefont {Zhang}},
  \bibinfo {author} {\bibfnamefont {B.}~\bibnamefont {Grabowski}}, \bibinfo
  {author} {\bibfnamefont {A.~V.}\ \bibnamefont {Shapeev}},\ and\ \bibinfo
  {author} {\bibfnamefont {A.~V.}\ \bibnamefont {Ruban}},\ }\href
  {https://doi.org/10.1103/physrevb.103.195428} {\bibfield  {journal} {\bibinfo
   {journal} {Physical Review B}\ }\textbf {\bibinfo {volume} {103}},\ \bibinfo
  {pages} {195428} (\bibinfo {year} {2021})}\BibitemShut {NoStop}%
\bibitem [{\citenamefont {Fultz}(2010)}]{ful10}%
  \BibitemOpen
  \bibfield  {author} {\bibinfo {author} {\bibfnamefont {B.}~\bibnamefont
  {Fultz}},\ }\href {https://doi.org/10.1016/j.pmatsci.2009.05.002} {\bibfield
  {journal} {\bibinfo  {journal} {Prog. Mat. Sci.}\ }\textbf {\bibinfo {volume}
  {55}},\ \bibinfo {pages} {247} (\bibinfo {year} {2010})}\BibitemShut
  {NoStop}%
\bibitem [{\citenamefont {Flubacher}\ \emph {et~al.}(1959)\citenamefont
  {Flubacher}, \citenamefont {Leadbetter},\ and\ \citenamefont
  {Morrison}}]{flu59}%
  \BibitemOpen
  \bibfield  {author} {\bibinfo {author} {\bibfnamefont {P.}~\bibnamefont
  {Flubacher}}, \bibinfo {author} {\bibfnamefont {A.~J.}\ \bibnamefont
  {Leadbetter}},\ and\ \bibinfo {author} {\bibfnamefont {J.~A.}\ \bibnamefont
  {Morrison}},\ }\href {https://doi.org/10.1080/14786435908233340} {\bibfield
  {journal} {\bibinfo  {journal} {Philos. Mag.}\ }\textbf {\bibinfo {volume}
  {4}},\ \bibinfo {pages} {273} (\bibinfo {year} {1959})},\ \bibinfo {note} {in
  Table 5, page 285 of this paper, "cal/g atom" should read
  "cal/mol".}\BibitemShut {Stop}%
\bibitem [{\citenamefont {Barin}(1995)}]{bar95}%
  \BibitemOpen
  \bibfield  {author} {\bibinfo {author} {\bibfnamefont {I.}~\bibnamefont
  {Barin}},\ }\href {https://doi.org/10.1002/9783527619825} {\emph {\bibinfo
  {title} {Thermochemical data of pure substances}}},\ \bibinfo {edition}
  {3rd}\ ed.\ (\bibinfo  {publisher} {VCH Verlagsgesellschaft mbH},\ \bibinfo
  {year} {1995})\BibitemShut {NoStop}%
\bibitem [{\citenamefont {Ashcroft}\ and\ \citenamefont
  {Mermin}(1976)}]{Ashcroft1976}%
  \BibitemOpen
  \bibfield  {author} {\bibinfo {author} {\bibfnamefont {N.~W.}\ \bibnamefont
  {Ashcroft}}\ and\ \bibinfo {author} {\bibfnamefont {N.~D.}\ \bibnamefont
  {Mermin}},\ }\href@noop {} {\emph {\bibinfo {title} {{S}olid {S}tate
  {P}hysics}}}\ (\bibinfo  {publisher} {Sauders College Publishing},\ \bibinfo
  {address} {New York},\ \bibinfo {year} {1976})\BibitemShut {NoStop}%
\bibitem [{\citenamefont {Modesti}\ \emph {et~al.}(2020)\citenamefont
  {Modesti}, \citenamefont {Sheverdyaeva}, \citenamefont {Moras}, \citenamefont
  {Carbone}, \citenamefont {Caputo}, \citenamefont {Marsi}, \citenamefont
  {Tosatti},\ and\ \citenamefont {Profeta}}]{mod20}%
  \BibitemOpen
  \bibfield  {author} {\bibinfo {author} {\bibfnamefont {S.}~\bibnamefont
  {Modesti}}, \bibinfo {author} {\bibfnamefont {P.~M.}\ \bibnamefont
  {Sheverdyaeva}}, \bibinfo {author} {\bibfnamefont {P.}~\bibnamefont {Moras}},
  \bibinfo {author} {\bibfnamefont {C.}~\bibnamefont {Carbone}}, \bibinfo
  {author} {\bibfnamefont {M.}~\bibnamefont {Caputo}}, \bibinfo {author}
  {\bibfnamefont {M.}~\bibnamefont {Marsi}}, \bibinfo {author} {\bibfnamefont
  {E.}~\bibnamefont {Tosatti}},\ and\ \bibinfo {author} {\bibfnamefont
  {G.}~\bibnamefont {Profeta}},\ }\href
  {https://doi.org/10.1103/PhysRevB.102.035429} {\bibfield  {journal} {\bibinfo
   {journal} {Phys. Rev. B}\ }\textbf {\bibinfo {volume} {102}},\ \bibinfo
  {pages} {035429} (\bibinfo {year} {2020})}\BibitemShut {NoStop}%
\bibitem [{\citenamefont {Sato}\ \emph {et~al.}(2000)\citenamefont {Sato},
  \citenamefont {Kitamura},\ and\ \citenamefont {Iwatsuki}}]{sat00}%
  \BibitemOpen
  \bibfield  {author} {\bibinfo {author} {\bibfnamefont {T.}~\bibnamefont
  {Sato}}, \bibinfo {author} {\bibfnamefont {S.}~\bibnamefont {Kitamura}},\
  and\ \bibinfo {author} {\bibfnamefont {M.}~\bibnamefont {Iwatsuki}},\ }\href
  {https://doi.org/10.1116/1.582283} {\bibfield  {journal} {\bibinfo  {journal}
  {J. Vac. Sci. Tech. A}\ }\textbf {\bibinfo {volume} {18}},\ \bibinfo {pages}
  {960} (\bibinfo {year} {2000})}\BibitemShut {NoStop}%
\bibitem [{\citenamefont {Chang}\ and\ \citenamefont {Wei}(2003)}]{cha03}%
  \BibitemOpen
  \bibfield  {author} {\bibinfo {author} {\bibfnamefont {C.~M.}\ \bibnamefont
  {Chang}}\ and\ \bibinfo {author} {\bibfnamefont {C.~M.}\ \bibnamefont
  {Wei}},\ }\href {https://doi.org/10.1103/PhysRevB.67.033309} {\bibfield
  {journal} {\bibinfo  {journal} {Phys. Rev. B}\ }\textbf {\bibinfo {volume}
  {67}},\ \bibinfo {pages} {033309} (\bibinfo {year} {2003})}\BibitemShut
  {NoStop}%
\bibitem [{\citenamefont {Cherepanov}\ and\ \citenamefont
  {Voigtl{\"a}nder}(2004)}]{che04}%
  \BibitemOpen
  \bibfield  {author} {\bibinfo {author} {\bibfnamefont {V.}~\bibnamefont
  {Cherepanov}}\ and\ \bibinfo {author} {\bibfnamefont {B.}~\bibnamefont
  {Voigtl{\"a}nder}},\ }\href {https://doi.org/10.1103/PhysRevB.69.125331}
  {\bibfield  {journal} {\bibinfo  {journal} {Phys. Rev. B}\ }\textbf {\bibinfo
  {volume} {69}},\ \bibinfo {pages} {125331} (\bibinfo {year}
  {2004})}\BibitemShut {NoStop}%
\end{thebibliography}

%

\end{document}